\documentclass[12pt]{iopart}
\usepackage{iopams}
\usepackage[nosort]{cite}
\usepackage[bookmarks=false]{hyperref}

\usepackage{bbding}
\usepackage{epsfig}
\usepackage{graphicx}
\usepackage{color}
\usepackage{dcolumn}
\usepackage{bm}
\usepackage{ulem}
\usepackage{lipsum}
\usepackage{amssymb}
\newcommand{\be}{\begin{equation}}
\newcommand{\ee}{\end{equation}}
\newcommand{\bea}{\begin{eqnarray}}
\newcommand{\eea}{\end{eqnarray}}

\begin{document}
\title[]{Transport through an AC driven impurity: Fano interference and bound states in the continuum }

\author{Sebasti\'an A. Reyes$^1$, Daniel Thuberg$^1$, Daniel P\'erez$^1$, 
Christoph Dauer$^2$, 
 and Sebastian Eggert$^{2}$}

\address{$^1$Instituto de F\'isica and Centro de Investigaci\'on en Nanotecnolog\'ia y Materiales Avanzados, Pontificia Universidad Cat\'olica de Chile, Casilla 306, Santiago 22, Chile}
\address{$^2$Physics Department and Research Center OPTIMAS, University of Kaiserslautern, D-67663 Kaiserslautern, Germany}

\begin{abstract}
Using the Floquet formalism we study transport through an AC-driven impurity in a tight binding chain. The results obtained are exact and valid for all frequencies and barrier amplitudes. At frequencies comparable to the bulk bandwidth we observe a breakdown of the transmission $T=0$ which is related to the phenomenon of Fano resonances associated to AC-driven bound states in the continuum. We also demonstrate that the location and width of these resonances can be modified by tuning the frequency and amplitude of the driving field. At high frequencies there is a close relation between the resonances and the phenomenon of coherent destruction of tunneling. { As the frequency is lowered 
no more resonances are possible 
below a critical value  
and the results approach a simple time average of the static transmission.}
\end{abstract}


\section{Introduction}

In recent years there has been remarkable progress in experimental techniques, making possible the fabrication of nanoscale quantum systems with a high degree of coherence and controllability. Systems containing just a few molecules are of particular interest since they are promising candidates for the realization of electronic components on the sub-silicon scale  such as
molecular electronics \cite{aviram1974molecular, balzani2006molecular, reed1997conductance, cui2001reproducible, nitzan2003electron, goser2004nanoelectronics, fagas2001electron, cuniberti2002fingerprints,shen2010electron, gutierrez2002theory}.
A fruitful line of research makes use of simplified tight-binding models that are able to capture the essential physics, gaining qualitative understanding of the transport mechanisms involved \cite{fagas2001electron, cuniberti2002fingerprints,shen2010electron, gutierrez2002theory}.
These models are also applicable when investigating other interesting systems such as quantum dot arrays \cite{van2002electron} or photonic materials \cite{bayindir2000tight, photonic2, bloch}.
Furthermore, in practical applications time-dependent effects such as  electromagnetic radiation or gate voltages can be used to manipulate the transport properties of nanodevices \cite{van2002electron,longhi2013floquet, ac-field, pump4}. In fact, photon assisted tunneling has been observed experimentally in a number of quantum resonant 
tunneling structures \cite{van2002electron,tsu1973tunneling} and high-speed switching devices could be designed based on this phenomenon. Periodically driven tunneling has been studied before by several authors \cite{shirley1965solution, li1999floquet, oberthaler, dressed-matter, harper}. 
Interestingly, recent works have highlighted the possibility of observing {\it AC-driven} bound states in the continuum (BIC's) in transport experiments \cite{longhi2013floquet,gonzalez2013bound,thuberg,fano}. First proposed by von Neumann and Wigner \cite{wigner}, BIC's consist of spatially localized states embedded in the spectrum of scattering states. The transmission and corresponding conductance of the device would show dynamical Fano resonances, revealing the existence of these exotic states. Furthermore, the energy of the BIC's can be changed by tuning the frequency of the driving field.

\begin{figure}
\begin{center}
\includegraphics[width=0.7\columnwidth]{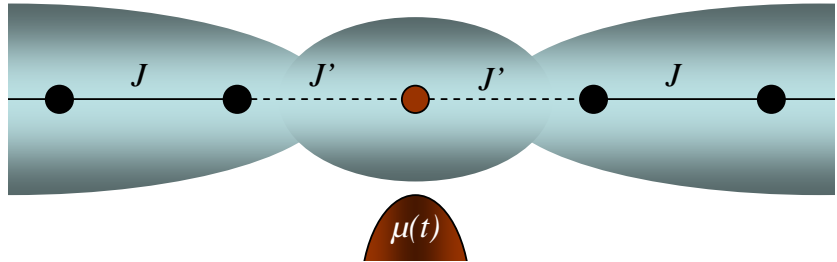}\\
\end{center}
\caption{Schematic setup for transport through a periodically driven quantum dot.}
\label{model-fig}
\end{figure}

{ In order to consider the effect of local driving on the transport along 
one direction, we study a tight binding chain which is periodically driven at one site.  
In realistic systems there are of course many parallel transport channels, but to describe
the scattering from a local periodic potential it is possible to consider 
each channel independently in form of a one-dimensional chain.  This model also allows to 
modify the tunneling amplitude $J'$ into the driven impurity site as depicted in Fig.~\ref{model-fig}.   
{ Typical experimental systems we have in mind are driven quantum dots, which are attached to leads or atomic quantum gases in tailored optical lattices with a time dependent impurity site. } 
Our results can be generalized to continuous systems by taking the lattice constant
of the tight binding model to zero.  { The continuum case is also interesting beyond impurity physics to describe the effective two-body interaction between atoms in a one-dimensional gas, which is subject to an oscillating Feshbach resonance \cite{hudson}.}
We obtain the exact transmission probability through the impurity based on the Floquet 
formalism \cite{floquet,book,grifoni}, which allows for a treatment beyond the perturbative regime. Our results reveal an array of phenomena that emerge for different tuning of the parameters.  
From previous works it is known that the driven impurity problem shows remarkable features 
such as tunability to  perfect transmission 
or to a  complete blockade at higher frequencies \cite{li1999floquet,thuberg,entin}.  
Such a 
{\it quantum resonance catastrophe} \cite{thuberg}
can even occur at infinitesimally small driving amplitude, 
if the frequency $\omega$ is slightly 
larger than the energy of the incoming particle relative to the upper or lower 
band edge.   
  It is also known that at very high 
frequencies a driven chemical potential can generally be mapped to a static effective
model involving Bessel functions \cite{della2007visualization,shaking}.
In this paper we now focus also on the effect of modified tunneling amplitudes
$J'$ as
well as {\it lower} frequencies, which have so far received little attention since 
an analytical averaging is not possible in that limit.  
We can show exactly that no resonances occur below a critical
frequency and how the transmission probability approaches the static limit.
}

In the following section we explain the model and formalism used in our calculations. Later, we present the results highlighting the main features found. In the last section we analyze our results and present some conclusions. \\

\section{The Model}
We consider excitations traveling along an infinite one-dimensional tight-binding chain with nearest neighbor hopping amplitude $J$. The central site is connected to the left and right sides of the chain by a modified coupling $J'$ and is periodically driven with angular frequency $\omega$ and amplitude $\mu$ as depicted in Fig.~\ref{model-fig}. The Hamiltonian of the system reads 
\be
\!\!\!\!\!\!\!\!\!\!\!\!\!\!\!\!\!\!\!\!\!\!\!\!\!\!\!\!\!\!\!\!\!H = -J \sum_{j\neq-1,0} (c_j^\dagger c_{j+1}{\phantom{\dagger}} + c_{j+1}^\dagger c_{j}{\phantom{\dagger}}) \ -J' \sum_{j=-1,0} (c_j^\dagger c_{j+1}{\phantom{\dagger}} + c_{j+1}^\dagger c_{j}{\phantom{\dagger}}) \ - \ \mu \cos (\omega t) c_0^\dagger c_{0}{\phantom{\dagger}},
\label{model}
\ee
where the operator $c_j^\dagger$ creates a particle on site $j$. This model captures the essential physics of an oscillating barrier connected to leads according 
to the possible experimental realizations mentioned above.  {
Since only independent particles in separate channels are considered, 
the following results apply equally
to bosonic or fermionic particles, so a simple integration over occupied states can be 
performed depending on the experimental situation.}

\begin{figure}[!t]
\begin{center}
\includegraphics[height=0.4\columnwidth]{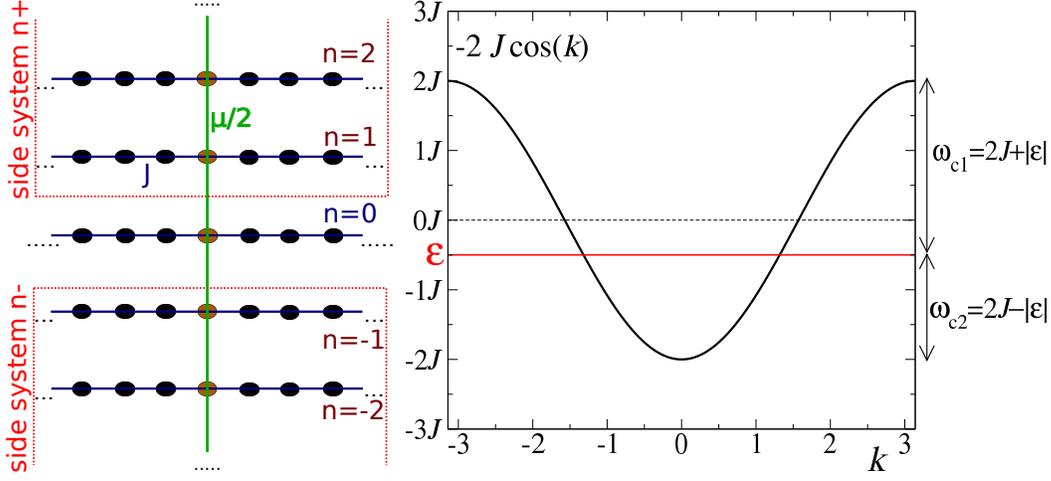}
\includegraphics[height=.4\columnwidth]{dispersion.eps}\\
\end{center}
\caption{Left: Sketch of the model mapped onto a set of static
coupled chains. The $n = 0$ chain is locally connected to two 
side-coupled systems of chains with a corresponding chemical potential
$n\omega$. Right: Dispersion relation for the $n=0$ chain. For frequencies below the critical value $\omega_{c1} = 2J + |\epsilon|$ there will be at least one second  unbound channel. }
\label{floquet-fig}
\end{figure}

In this paper we calculate the transmission probability $T$ for incoming particles with momentum $k_0$ and corresponding energy $\epsilon=-2J\cos k_0$.   
To do this we will find steady-state solutions to the Schr\"odinger equation (using $\hbar=1$)
\be
(H(t) -i \partial_t) \left|\Psi(t)\right\rangle = 0. \label{Eq:Schrodinger}
\ee
Due to the periodicity of the Hamiltonian $H(t) = H(t+2 \pi/\omega)$, we can apply the Floquet formalism \cite{floquet,book,grifoni} to find steady-state solutions in terms of the so-called Floquet states $\left|\Psi(t)\right>=e^{-i\epsilon t}\left|\Phi(t)\right>$, where $\left|\Phi(t)\right> = \left|\Phi(t+2 \pi/\omega)\right>$ are the Floquet modes and $\epsilon$ its quasienergy corresponding to the eigenvalue equation 
\be
(H(t) -i \partial_t) \left|\Phi(t)\right> = \epsilon \left|\Phi(t)\right>. \label{Eq:eigenvalue}
\ee
Using a spectral decomposition for the Floquet modes
\be
|\Phi(t)\rangle = \sum_{n=-\infty}^{\infty}e^{-i n \omega t} |\Phi_{n}\rangle, \label{Eq:spectral}
\ee
we can obtain the eigenvalue equation for a Hamiltonian of the form $H(t)=H_0+2H_1\cos(\omega t)$:
\be
H_0|\Phi_n\rangle + H_1(|\Phi_{n+1}\rangle+|\Phi_{n-1}\rangle) = (\epsilon +n\omega)|\Phi_{n}\rangle
\ee
Now consider the general form of one of the time independent components of the spectral decomposition
\be
|\Phi_{n}\rangle = \sum_j \phi_{j,n} c_j^\dagger |0\rangle, \label{Eq:staten}
\ee
where $|0\rangle$ is the vacuum state. Inserting (\ref{Eq:staten}) into the eigenvalue equation (\ref{Eq:eigenvalue}) results in the following recursive relations for the amplitudes $\phi_{j,n}$,
\bea
-J (\phi_{j-1,n}+\phi_{j+1,n}) &=& (\epsilon+n\omega) \phi_{j,n}  ~~{\rm for }~  |j| > 1 \nonumber \\
-J \phi_{\pm 2,n} -J' \phi_{0,n} &=& (\epsilon+n\omega) \phi_{\pm 1,n}   \label{condnj} \\
-J' (\phi_{-1,n}+\phi_{1,n})-\frac{\mu}{2}(\phi_{0,n+1}+\phi_{0,n-1}) &=& (\epsilon+n\omega) \phi_{0,n} . \nonumber
\eea
These may be interpreted as a mapping of our original time-periodic problem into a time independent one with an infinite number of tight-binding chains labeled by $n$, each with overall chemical potential $n\omega$. Each chain is associated with a Fourier mode and is coupled to its nearest neighbors by a hopping term $\mu/2$ at the central site $j=0$ (see Fig.~\ref{floquet-fig}).

\section{Transmission Coefficient}
We now proceed to calculate the exact transmission coefficient $T$ for an incoming wave with wavenumber $k_0$ and energy $\epsilon = -2J\cos k_0$, corresponding to $n=0$.  Notice from 
equation~(\ref{condnj}) that the bulk of each chain supports solutions of the form $e^{ ik_n j}$
with wavenumbers given by $-2J\cos k_n = \epsilon + n\omega$. 
When $|\epsilon+n\omega|<2J$, $k_n$ is real and the corresponding plane wave solutions are delocalized over the entire chain (unbound channels). We will assume that $n=0$ is an unbound channel and contains the incoming wave of unit amplitude, a reflected part of amplitude $r_0$ and a transmitted wave with amplitude $t_0$. For all unbound channels plane waves with wavenumbers $ k_n $ can emerge from the impurity as additional transmitted and reflected parts. 
For chains with $|\epsilon+ n\omega|>2J$, $k_n$ is imaginary and the solutions decay exponentially around the impurity (bound channels). Bound channels with negative (positive) energy have $k_n=i\kappa_n$ ($k_n=i\kappa_n+\pi$) where $\kappa_n$ is real. The corresponding energies are $\epsilon + n\omega= \mp 2J\cosh \kappa_n$. 

Thus, for our setup with one incoming wave the solutions for all channels $n$ can be written as
\be 
|\Phi_n\rangle = \sum_{j<0} (\delta_{n,0} e^{i k_0 j} + r_n e^{-i k_n j}) c^{\dagger}_j | 0 \rangle + \sum_{j>0} t_n e^{i k_n j} c^{\dagger}_j | 0 \rangle + E_n \frac{J}{J'} c^{\dagger}_0 | 0 \rangle. \label{Ansatz}
\ee
Using equation~(\ref{condnj}) it is easy to check that
\be
E_n=t_n=r_n+\delta_{n,0}, \label{rtn}
\ee
capturing the inversion symmetry of the lattice with respect to $j=0$. We can then derive the amplitudes at each site for the Floquet modes: 
\be 
\phi_{j,n} = E_n e^{ik_n|j|} + \delta_{j,0} E_n \left(\frac{J}{J'} - 1\right) + \delta_{n,0} 2i\sin (k_0 j) \Theta(-j)\label{Ansatz2}.\\
\ee
Inserting this result back into equation~(\ref{condnj}) we obtain a recursive relation for the coefficients $E_n$:
\be
E_{n+1} + E_{n-1} = \frac{4J}{\mu}\left[E_n(\beta e^{ik_n} - i\sin k_n)+i\delta_{n,0}(1-\beta)\sin k_n \right]  \label{rec_c}, 
\ee
where we have defined $\beta = 1-(J'/J)^2$ { to parametrize 
the inhomogeneity in hopping}. As explained in detail below, this recursive relation will be central to all of our calculations of the transmission coefficient.

The total transmission probability consists of the ratio between the transmitted and incoming current. In the present model all unbound channels contribute with their individual transmission, resulting in 
\be 
T = \sum_n T_n =  \sum_{n} |t_n|^2 \frac{\sin k_n}{\sin k_0}, \label{Eq_T_1}
\ee
where the sum is over all the extended states. On the other hand, current conservation requires that the incoming current is equal to the outgoing current:
\be 
 \sin k_0 = \sum_{n} \left(|r_n|^2 + |t_n|^2 \right) \sin k_n. \label{Eq_Current}
\ee
Now, we can combine equations (\ref{rtn}), (\ref{Eq_T_1}) and (\ref{Eq_Current}) to obtain (see \ref{App_A}):
\be 
T = {\rm Re} E_0 = {\rm Re}\left[\frac{u_{k_0} (1 - \beta)}{u_{k_0} (1-\beta)-i\beta\epsilon - \frac{i\mu}{2}\left(\frac{E_1}{E_0}+\frac{E_{-1}}{E_0}\right)}\right]\label{eq_trans},
\ee 
where $u_{k_0}=2J\sin k_0$ is the velocity of the incoming particle. Thus, to calculate $T$ we need to determine the ratios $E_{\pm1}/E_0$,  which are completely fixed by requiring convergence of the above recurrence relations in equation (\ref{rec_c}) as $n\to\pm\infty$. For a given set of physical parameters, these ratios can be obtained numerically in a very fast and efficient manner.

It is important to consider the zero transmission resonances that occur when ${\rm Re} E_0=0$. By imposing this condition upon our recursive relations we can derive an approximation for small $\mu$ which indicates the location of the resonances (see \ref{App_B}):
\be
\!\!\!\!\!\!\!\!\!\!\!\!\!\!\!\!\!\!\!\!\!\!\!\!\!\!\!\!\!\!\!\!\!\!\!\!\!\!\!\!\!\!\!\!\mu^2\approx 4\left[\mp(1-\beta)\sqrt{(\epsilon\pm2\omega)^2-4J^2}-\beta(\epsilon\pm2\omega)\right]\left[\mp(1-\beta)\sqrt{(\epsilon\pm\omega)^2-4J^2}-\beta(\epsilon\pm\omega)\right]  \label{res_app}.
\ee
For $J'\geq J$ ($\beta\leq0$) this equation has a solution for $\mu\rightarrow 0$, if
 the frequency $\omega_r$ at which the resonances appear in the limit of vanishing barrier amplitude are given by 
\be
\omega_r= \frac{2J(1-\beta)}{\sqrt{1-2\beta}} \mp \epsilon \label{res_omega}.
\ee
In the opposite case, when there is a weaker coupling to the impurity ($0 < \beta < 1$), equation~(\ref{res_app}) does not have a solution for $\mu=0$. Here the resonances emerge at frequencies $\omega_r=2J\mp\epsilon$ independent of $\beta$, but at finite barrier amplitudes given by \be
\mu_{r\pm}^2=8\beta J\left[(1-\beta)\sqrt{(4J\mp\epsilon)^2-4J^2}+\beta(4J\mp\epsilon)\right]\label{res_mu}
\ee

According to equation~(\ref{Eq_T_1}) the condition for resonance $T=0$ also requires that $t_n=0$ for all
unbound { channels}, since $\sin k_n > 0$ within the band.  This has an interesting consequence if
$\omega < \omega_{c1}=2J+|\epsilon|$, i.e.~when there are at least two unbound { channels}
(see Fig.~\ref{floquet-fig}). For example assuming $\epsilon>0$, then for resonance 
the two unbound { channels},  at $n=0$ and $n=-1$ must obey
$t_0=t_{-1}=0$ and 
according to equation~(\ref{rec_c}) for $n=0$, this implies 
\be E_1=t_1 = -i \frac{4 J}{\mu}(1-\beta) \sin k_0.
\ee
  Therefore, $t_1$ cannot be zero, except for trivial cases, such as 
$k_0=0$ or $J'=0$.  But since $t_n=0$ for all
unbound { channels}, this implies that the channel for $n=1$ must lie
outside the band in case
of resonance, i.e.~$\epsilon_1 = \epsilon + \omega > 2 J$.  
The same reasoning applies for $\epsilon < 0$ and unbound { channels} for 
$n=0$ and $n=1$, giving the condition $\epsilon_{-1} = \epsilon - \omega < -2 J$, 
so that there can never be more than two unbound channels if $T=0$.
In summary, the condition for zero transmission therefore requires that 
$\omega > \omega_{c2} =2J-|\epsilon|$, which provides a lower cutoff for the observation of possible  
resonances.

In the following section we will present the exact transmission values for our system and indicate approximations for various regimes.

\section{Results and analysis}
{ 
\subsection{Homogeneous chains $J'=J$}
The homogeneous chain has been explored before, but only in the 
 high frequency region \cite{thuberg}.
Resonances occur 
even for small values of $\mu$ if the frequency is tuned just above
the critical frequencies $\omega_{c1/2}$.  In particular, for small $\mu$ the 
location of zero transmission 
can be approximated as \cite{thuberg} 
\be
\omega_{r\pm} \approx 2J \pm \epsilon+{\mu^4\over 64 J\left([4 J\pm\epsilon]^2-4J^2\right)}
\ee
as derived in equation~(\ref{res_app}) (see also \ref{App_B}).
This behavior is illustrated in Fig.~\ref{results_2} (top) for $\epsilon=-0.5J$ as a
function of $\mu$ and $\omega$ by red dashed lines.  
Comparison with the blue line marking the exact location of the resonance shows that this is a good approximation up to $\mu\approx 3J$.
As the amplitude is increased these zero transmission resonances become wider and move to higher frequencies. 

\begin{figure}[!t]
\begin{center}
\includegraphics[width=0.7\columnwidth]{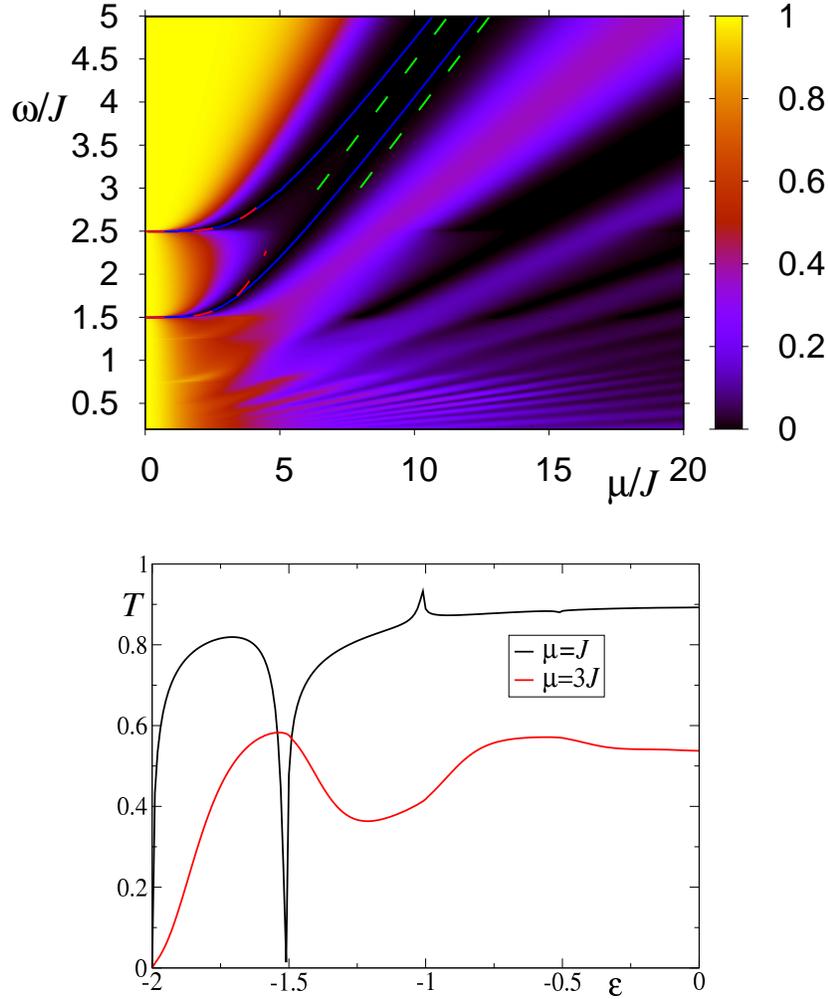}\\~\\
\includegraphics[width=0.5\columnwidth]{omega0.5-Jp1.eps}
\end{center}
\caption{Exact results for transmission probability $T$ {\it Top}:  
for an incoming wave of energy $\epsilon = -0.5J$  as a function 
of amplitude $\mu$ and frequency $\omega$. The dashed green line 
corresponds to high frequency  approximation according to equation~(\ref{approx_high_freq}), 
while the dashed red line corresponds to equation~(\ref{approx_eps}).           
{\it Bottom:} 
as a function of energy $\epsilon$ of the incoming particle for $\omega=0.5J$ and different values of the barrier amplitude $\mu$ }
\label{results_2}
\end{figure}

Turning our attention to lower frequencies 
in Fig.~\ref{results_2} (top), 
below $\omega_{c2} = 2J-\epsilon = 1.5 J$ there are no zero-transmission resonances
as predicted, but
it is remarkable that 
their remnants are still visible in the form of  oscillating structures as a function of 
$\mu$ and $\omega$ in the low frequency range.  Fixing the frequency to a relatively 
small value $\omega=0.5J$ the resonance near $\epsilon = \pm (2J-\omega)$ is  robust for 
small values of $\mu$ as shown in Fig. \ref{results_2} (bottom).  
The location of this resonance can again be approximated for small $\mu$ 
using  equation~(\ref{res_app}) for $\beta=0$
\be
\epsilon_r \approx \pm\left(2J -\omega + \frac{\mu^4}{ 64 J \omega (4J + \omega)}\right) \label{approx_eps},
\ee
which implies that with increasing $\mu$ the energy of the resonance is
moved towards the band edge for small frequencies $\omega<2J$ while
it moves towards the center of the band for $\omega>2J$.
Accordingly, no resonance can be found for $\mu=3J$ and $\omega=0.5J$ in 
Fig.~\ref{results_2} (bottom), 
since it has been pushed outside the band.}

%

%

It is interesting to note that the observed resonances can be directly related to the phenomenon of Fano interference, which is known to affect the transmission of waves in discrete arrays \cite{miroshnichenko2005engineering}. This interference effect occurs in systems where one or more discrete energy levels interact with the continuum spectrum, leading to resonances located precisely at the energy of these states. As pointed out above, our model of a discrete chain with an AC-barrier can be mapped onto a set of coupled chains. Thus, in this case the continuum spectrum corresponds to the $n=0$ chain, whereas the discrete states are the bound states of the ``side-attached'' system composed by the virtual chains associated to higher frequency modes ($n\neq0$, see Fig. \ref{floquet-fig}). These states can also be interpreted as the AC-driven bound states in the continuum explained above. We have performed an independent numerical calculation to obtain the values of the aforementioned discrete energies and 
confirmed 
that they indeed coincide with the location of the transmission resonances calculated before.  { For larger amplitude $\mu$ additional bound states of the side coupled system can also be pushed into the band, which show up as higher order resonances as can be seen in Fig.~\ref{results_2} e.g.~for $\omega>\omega_{c2}$ and $\mu \gtrsim 7.5J$.}   Notice that the eigenenergies of the upper 
and lower ``side-system'' only differ by their sign, which is consistent with the fact that our recursive relations already were invariant under inversion of the incoming particle energy $\epsilon \leftrightarrow -\epsilon$. 

\begin{figure}[!t]
\begin{center}
\includegraphics[width=0.5\columnwidth,angle=-90]{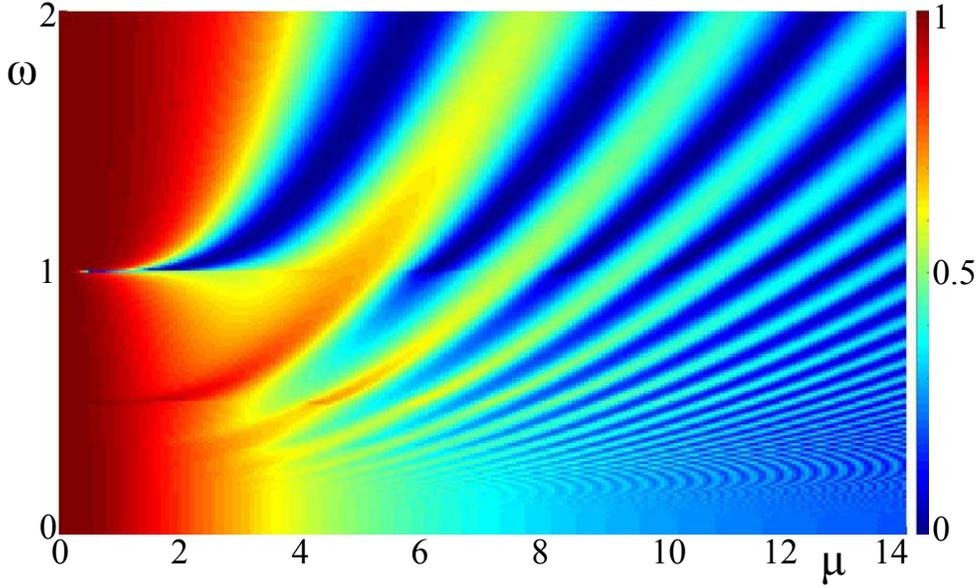}
\end{center}
\caption{Transmission coefficient in the limit of vanishing lattice spacing for 
$\epsilon'=1$ in units of $1/2m$, 
i.e. for a quadratic dispersion relation with a local driving in form of a delta-function}
\label{cont-limit}
\end{figure}
\subsection{Continuum limit}
One advantage of the tight-binding model is that it can readily be generalized to a continuum system by taking the lattice spacing to zero.  The effective mass of the dispersion relation
is then determined  by the second order expansion of the cosine dispersion relation
$\epsilon'= 2J -2 J \cos k a  \to J a^2 k^2 = k^2/2 m$, i.e.~the hopping $J$ is taken to be infinite at the same time, so that $J a^2=1/2m$ is finite.  Here we have shifted
$\epsilon'= \epsilon + 2J$, so that 
$\epsilon' > 0$
corresponds to unbound { channels} and $\epsilon' < 0$ are bound { channels}.   
Using the analogous approach as used for equation~(\ref{rec_c})
above we now obtain for the recursive relation
\be
E_{n+1} + E_{n-1} = -i\frac{2}{m \mu}k_n\left( E_n-\delta_{n,0}  \right)  \label{rec_c2}, 
\ee
where $k_n= \sqrt{2m (\epsilon+n\omega)}$.
The resulting transmission coefficient is plotted in Fig.~\ref{cont-limit} for an 
incoming energy of $\epsilon'= 1$ in units of $1/2m$.  Note, that the continuum problem has become scale 
invariant, so that a rescaling of $\epsilon'$ by a factor $c$ leads to the same results
as a function of 
$c \omega$ and $\sqrt{c} \mu$, 
since there is no longer a finite bandwidth to dictate an intrinsic energy scale.
Accordingly, Fig.~\ref{cont-limit} contains the complete information about the energy
dependence as well.
The lower critical frequency is now given by $\omega_{c2} = \epsilon'$ and the 
higher critical frequency $\omega_{c1}$ is pushed beyond bounds since 
there is no upper band edge.  Resonances appear only for frequencies $\omega > \epsilon'$, 
which show qualitatively  a similar structure as for the tight-binding model.  However, 
as we will see below an analytical high-frequency approximation 
in terms of Bessel functions exists for the tight binding model, 
which is not possible for the continuous system.

\subsection{Inhomogeneous coupling to the impurity site $J'\neq J$}
Let us now go back to the tight binding model in order to 
explore what happens when we consider a different coupling to the impurity site ($J'\neq J$) corresponding to  physical realizations where the coupling 
to the driven impurity may also differ from the ones along the chain. 

\begin{figure}[!t]
\begin{center}
\includegraphics[width=0.7\columnwidth]{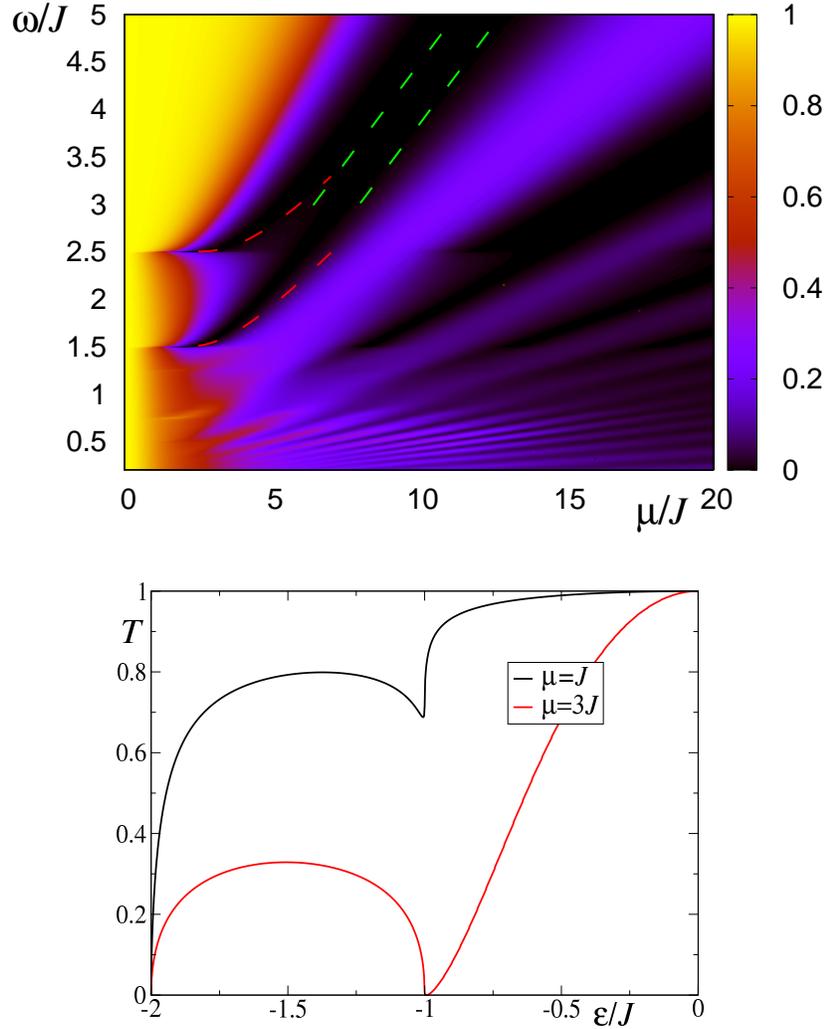}\\~\\
\includegraphics[width=0.5\columnwidth]{omega3-Jp09.eps}
\end{center}
\caption{Transmission coefficient for modified coupling $J'=0.9J$ {\it Top:}  
for a fixed energy $\epsilon=-0.5J$ as a function of $\mu$ and $\omega$. 
The dashed lines depict the analytic approximations for the location of the resonances at high frequencies (green), and close to where they emerge (red) from equation~(\ref{res_app})). 
{\it Bottom:} for fixed frequency $\omega=3J$.}          
\label{results_3}
\end{figure}

\begin{figure}[!t]
\begin{center}
\includegraphics[width=0.7\columnwidth]{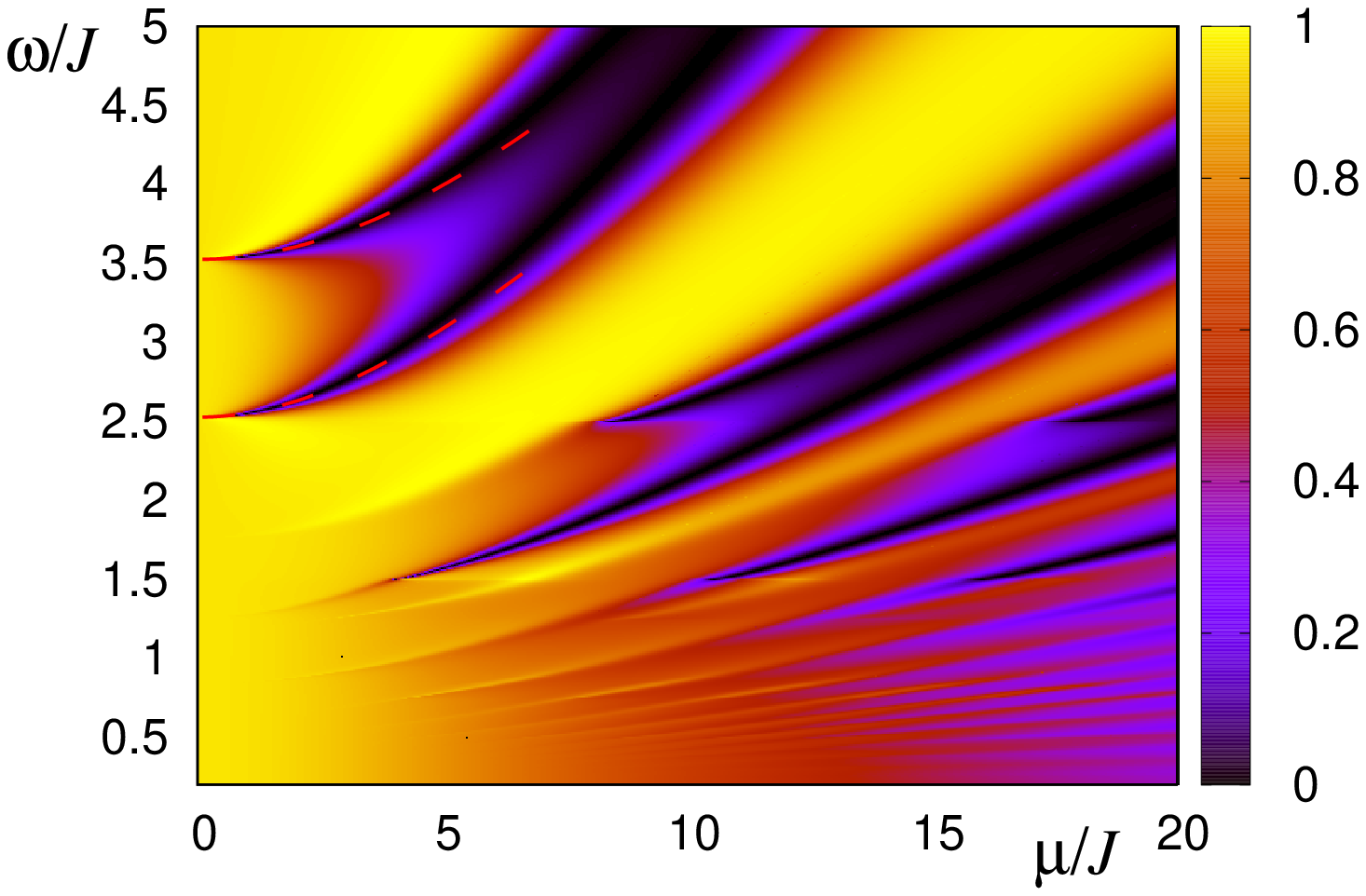}\\~\\
\includegraphics[width=0.5\columnwidth]{omega3-Jp2.eps}
\end{center}
\caption{Transmission coefficient for modified coupling $J'=2J$ {\it Top:}
for a fixed energy $\epsilon=-0.5J$ as a function of $\mu$ and $\omega$.
The red dashed lines depict the analytic approximations for the location of the resonances 
close to where they emerge from equation~(\ref{res_app})).
{\it Bottom:} for fixed frequency $\omega=3J$.}
\label{results_3-2}
\end{figure}

In Fig.~\ref{results_3} we show results for the transmission for a weaker hopping 
amplitude $J'=0.9J$. 
While there are similarities to the homogeneous case discussed above, the resonances
now start at a finite value of $\mu>0$ according to equation~(\ref{res_mu}).
As can clearly be seen in the bottom part of Fig.~\ref{results_3} for $\omega=3J$
accordingly there is a dip but no resonance for $\mu=J$, while
the resonance at $\mu=3J$ is relatively sharp.

As shown in Fig.~\ref{results_3-2} (top) 
for larger hopping amplitude $J'= 2J$
we observe a general increase in transmission across the parameter space 
and the resonances move to higher frequencies
for a given energy of the incoming wave following equation~(\ref{res_omega}).
Correspondingly, the resonances are also displaced in energy 
at fixed frequency $\omega=3J$ as shown in Fig.~\ref{results_3-2} (bottom).

These features can be interpreted nicely by using the Fano resonances due to the ``side-attached'' systems explained above and depicted in Fig. \ref{floquet-fig}. 
{ It turns out that the two resonances we observe in Figs.~\ref{results_3} and \ref{results_3-2} are associated with a bound state of the upper (lower) side system which is strongly localized on the $n=1$ ($n=-1$) chain. Let us consider $|J'-J|\ll 1$ such that each side system can be thought of as a set of coupled homogeneous chains ($J'=J$) plus a small perturbation. A simple first order calculation for the corresponding energy shift of the relevant bound states gives
\be
\delta\epsilon_{B_\pm} = -(J'-J)\sum_{n\gtrless 0} Re(\phi_{0,n}^{B_\pm *}\phi_{1,n}^{B_\pm }+\phi_{0,n}^{B_\pm *}\phi_{-1,n}^{B_\pm }),
\ee
where $B_+$ ($B_-$) denotes the bound state corresponding to the upper (lower) side system. It turns out that for the $n>0$ ($n<0$) side the amplitudes $\phi_{1,n}^{B}$ and $\phi_{-1,n}^{B}$ have the same (opposite) sign as $\phi_{0,n}^{B}$. 
Thus, as $J'$ becomes larger than $J$ the resonance corresponding to the ``$+$'' side (``$-$'' side) tends to decrease (increase) its energy and viceversa. We emphasize that for certain parameter combinations the BIC's energy shift may push it back into the continuum of states of the side system or away from energy band of the incoming particle. In either case this makes the resonance disappear. } 

It should be mentioned that the same formalism can be applied for an asymmetric 
tunneling junction with unequal tunneling amplitudes to the impurity site
$J'_{left}\neq J'_{right}$. In this case possible location of the resonances  
are again given by equations~(\ref{res_app})-(\ref{res_mu}) with $\beta= 1- (J'^2_{right} + J'^2_{left})/2J^2$.


\section{High Frequency}

In the high frequency regime ($\omega\gg J$) it is interesting to note that, as the amplitude of the barrier is increased for fixed frequency $\omega$, the transmission undergoes oscillations that reach $T=0$ for certain values of $\mu$ as shown in Figs.~\ref{results_2}, \ref{results_3} and \ref{results_3-2}. Moreover, the density plots suggest that for higher frequencies the transmission probability depends only on the ratio $\mu/\omega$.

How these features are related to our calculation can be explained in detail if we notice that the coefficient at the left of equation (\ref{rec_c}) becomes linear in $n$ and independent of $\beta$ when $\omega\gg J$ \cite{thuberg}. Thus, for high frequencies the recursive relation (\ref{rec_c}) is satisfied by Bessel functions of the first kind, i.e. $E_{n\lessgtr 0} \sim \mathcal{J}_{n \mp \epsilon/\omega}(\mu/\omega)$. Inserting this relation in equation~(\ref{eq_trans}) it is then possible to obtain an approximate analytical expression for this regime 
\be
T(\epsilon) \approx \frac{u_{k_0}^2(1 - \beta)^2}{u_{k_0}^2(1 - \beta)^2 + \left[\beta\epsilon+\frac{\mu}{2} \left(\frac{\mathcal{J}_{1 - \epsilon/ \omega}(\mu/ \omega)}{\mathcal{J}_{-\epsilon/ \omega}(\mu/ \omega)} - \frac{\mathcal{J}_{1 + \epsilon/ \omega}(\mu/ \omega)}{\mathcal{J}_{\epsilon/ \omega}(\mu/ \omega)}\right)\right]^2}\label{approx_high_freq}.
\ee
The points of total reflection are clearly identified by the zeros of the Bessel functions $\mathcal{J}_{\pm \epsilon/\omega}(\mu/\omega)$ and are marked by green dashed lines in Figs.~\ref{results_2} and \ref{results_3}. 
It has been shown before that the exact transmission compares well with the high frequency
approximation for $\omega \gtrsim 10J$ \cite{thuberg}.

For even higher frequencies we can get a good approximation by expanding the denominator in equation (\ref{approx_high_freq}) to lowest order in $\epsilon/\omega$. Both resonances merge on the zeros of $\mathcal{J}_{0}(\mu/\omega)$ and the transmission probability becomes
\be
T \approx \frac{u_{k_0}^2}{u_{k_0}^2 + \epsilon^2 \left[\frac{1 - \left(J'\mathcal{J}_0(\mu/ \omega)/J\right)^2}{\left(J' \mathcal{J}_0(\mu/ \omega)/J\right)^2}\right]^2}, \label{approx_high_freq_2}
\ee
This result is directly related to the phenomenon of coherent destruction if tunneling (CDT) which states that at high frequencies the effect of the oscillating impurity is averaged out, resulting in an effective tunneling to the impurity site given by $J_{\mbox{\tiny eff}}=J \mathcal{J}_0(\mu/\omega)$ \cite{della2007visualization}. Thus, in this regime the situation can be viewed as a time independent double barrier problem, where the middle region is the impurity site and there is an associated resonant tunneling ($T=1$) for $\epsilon=0$. In fact, the transmission for this effective static model can be calculated exactly, giving the result obtained in equation~(\ref{approx_high_freq_2}).
{ In other words, in this context we can view the phenomenon of CDT as a destructive Fano interference process in the regime where the width of the associated $T=0$ resonance diverges, blocking the whole energy spectrum. }
Note, however, that the averaging used to derive the CDT is a further 
simplification compared to our analytical description in terms of 
Bessel functions with fractional index in equation~(\ref{approx_high_freq}).

\section{Low Frequency}

At very low frequencies $\omega\ll \omega_{c2}$, { when the change in $\epsilon_n=\epsilon +n\omega$ 
becomes negligible in each recursive iteration 
it is reasonable to use approximately the same particle velocity in equation~(\ref{rec_c}), i.e. $u_{k_n} \approx u_{k_0}=2J \sin k_0$. 
In this limit we can derive an approximate analytic expression for the transmission.}
Let us consider a product ansatz for the coefficients of the type $E_n=E_0 \lambda^{\pm n}$ for $n\gtrless0$. We find from equation~(\ref{rec_c}) that
\be
\lambda^{2}-\frac{4J}{\mu}(\beta e^{ik_0} - i\sin k_0)\lambda+1 = 0,
\ee
which is solved by
\be
\lambda_{\pm}=\frac{2J}{\mu}(\beta e^{ik_0} - i\sin k_0)\pm\sqrt{\left(\frac{2J}{\mu}(\beta e^{ik_0} - i\sin k_0)\right)^2-1},
\ee
where $\lambda_+=1/\lambda_-$.  
Choosing $\lambda$ such that $|\lambda|<1$ and taking into account the finite norm condition that requires $E_n \to 0$ when $n\to\infty$, in good approximation for low frequencies we have
\be
\frac{E_1}{E_{0}}\approx \frac{E_{-1}}{E_{0}} \approx\frac{2J}{\mu}(\beta e^{ik_0} - i\sin k_0)\pm\sqrt{\left(\frac{2J}{\mu}(\beta e^{ik_0} - i\sin k_0)\right)^2-1}.
\ee
Finally, inserting these results into equation~(\ref{eq_trans}) we obtain an analytic result for low frequencies
\be
\!\!\!\!\!\!\!\!\!\!\!\!\!\!\!\!\!\!\!\!\!\!\!\!\!T_{\omega \gtrsim 0}\approx\frac{u_{k_0}}{2}\left|\frac{1}{\sqrt{\left(u_{k_0}-i\left(\frac{J^2}{J'^2}-1\right)\epsilon\right)^2+\mu^2\frac{J^4}{J'^4}}}+\frac{1}{\sqrt{\left(u_{k_0}+i\left(\frac{J^2}{J'^2}-1\right)\epsilon\right)^2+\mu^2\frac{J^4}{J'^4}}}\right|\label{trans_low_freq}
\ee

We now want to compare our result for $\omega\rightarrow 0$ with the one obtained from averaging a static barrier of oscillating amplitude. Let us consider then the Hamiltonian
\be
\!\!\!\!\!\!\!\!\!\!\!\!\!\!\!\!\!\!\!\!\!\!\!\!H_{static} = -J \sum_{i\neq-1,0} (c_i^\dagger c_{i+1}{\phantom{\dagger}} + c_{i+1}^\dagger c_{i}{\phantom{\dagger}}) \ -J' \sum_{i=-1,0} (c_i^\dagger c_{i+1}{\phantom{\dagger}} + c_{i+1}^\dagger c_{i}{\phantom{\dagger}}) \ - \ \mu c_0^\dagger c_{0}{\phantom{\dagger}}.
\label{model_static}
\ee
where $\mu$ is the constant magnitude of the barrier. 
Using the general ansatz equation~(\ref{Ansatz}) for only one conducting channel $n=0$ we get
\be
2E_{0} = \frac{4J}{\mu}\left[E_0(\beta e^{ik_0} - i\sin k_0)-i(1+\beta)\sin k_0 \right].
\ee
Thus, the equation for the static case looks just as the one for the dynamic situation, however since there are no neighbor chains to couple to, it couples with itself. Simple algebra yields
\be
T_{static} = |t|^2 = |E_0|^2 = \frac{u_k^2}{u_k^2 + \left(\mu-\frac{J^2}{J'^2}\epsilon\right)^2}.
\ee
Now, if we consider a very slow oscillating barrier ($\omega\rightarrow 0$) we need to average the result for the static transmission over a whole period $2\pi/\omega$:
\be
T_{\omega \rightarrow 0} = \frac{\omega}{2\pi} \int_0^{2\pi/\omega} \frac{u_k^2}{u_k^2 + \left(\mu\cos \omega t-\epsilon{J^2}/{J'^2}\right)^2}.
\ee
This definite integral can be calculated exactly, reproducing the result obtained above in equation~(\ref{trans_low_freq}).

\begin{figure}[!t]
\begin{center}
\includegraphics[width=0.6\columnwidth]{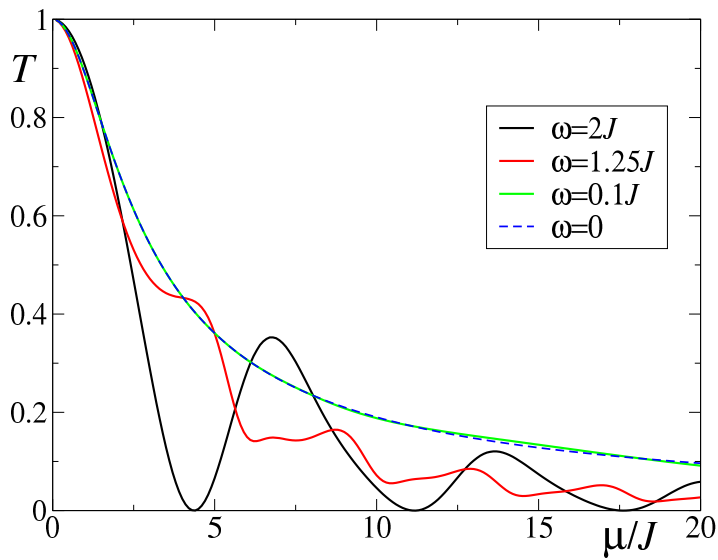}\\
 \end{center}
\caption{Transmission probability as a function of $\mu$ for different frequencies and
$\epsilon=-0.5$ and $J=J'$.}
 \label{results_6}
\end{figure}

{
Fig.~\ref{results_6} illustrates how the low frequency limit is approached
for the case of $\epsilon=-0.5$.}

\section{Conclusions}
Based on the Floquet formalism we have developed a framework to study the transmission probability across a time dependent impurity. We have included in our analysis the possibility of links to the impurity that are different from those in the bulk, as this may be the case in  potential experimental setups. 
The transmission profile is strongly influenced by the appearance of resonances that result in zero transmission for well defined energies of the incoming wave. {
Analysis for the inhomogeneous chain shows the important effects that a different coupling $J'\neq J$ to the impurity site may have. In particular, we observe how this parameter modifies the location of the aforementioned resonances, making them disappear for small barrier amplitudes when $J'<J$. We also showed that our calculation can be used to study the continuum limit where the results naturally become scale invariant. { In general, our results reveal a wide range of tuning possibilities for the location and shape of the resonances that can be modified by changing the amplitude and frequency of the perturbation.}

In the regimes of low and high frequency $\omega$ we derived closed analytical formulas for the transmission that are very good approximations to the exact results. A phenomenon closely related to CDT is present for very fast driving, 
where the impurity completely blocks the transmission for all energies at specific values of $\omega/\mu$. 
On the other hand, for driving much slower than $J$ our formalism reproduces the expected time average of the static case.
}

We demonstrate that the observed phenomena related to zero transmission resonances have a clear interpretation in terms of destructive Fano interference that appears when a linear chain is locally coupled to a system with a discrete set of states \cite{miroshnichenko2005engineering}. 
In the present case, although there is no real structure attached to the chain, due to the periodic perturbation the problem can be mapped into one where the conduction channel is coupled at the impurity site to an infinite set of 
``virtual'' chains (cf.~equation~(\ref{condnj})).


\ack
C.D. and S.E. are thankful for discussions with Axel Pelster.
This research was financially supported by CONICYT Grant No. 63140250 and the German Research Foundation (DFG) via the SFB Transregio 185.

\appendix 

\section{Transmission}\label{App_A}
We will show here the derivation leading to the transmission equation (\ref{eq_trans}). First, consider the current operator on each lattice site defined by
\be 
\hat{j}_m = - iJ (c_m^{\dagger} c_{m+1} - c_{m+1}^{\dagger} c_m).
\ee
We calculate the expectation value of the current on the negative $\langle \Phi_-|\hat{j}_m|\Phi_- \rangle$  as well as the positive side $\langle \Phi_+|\hat{j}_m|\Phi_+ \rangle$ of the chain, with
\be 
|\Phi_- \rangle = \sum_{j<0} \sum_{n} (e^{i k_0 j} + r_n e^{-i k_n j}) c^{\dagger}_j | 0 \rangle 
\ee
and
\be 
|\Phi_+ \rangle = \sum_{j>0} \sum_{n}  t_n e^{i k_n j}  c^{\dagger}_j | 0 \rangle, 
\ee
where $n$ labels the conducting (unbound) channels. The total current of each side then reads
\be 
J_- = \sum_{j<0} \langle \Phi_- |\hat{j}_m| \Phi_- \rangle = 2 J [\sin(k_0) - \sum_{n}  |r_n|^2 \sin(k_n)]
\ee
and
\be 
J_+ = \sum_{j>0} \langle \Phi_+ |\hat{j}_m| \Phi_+ \rangle = 2 J \sum_{n}  |t_n|^2 \sin(k_n) 
\ee
respectively. We now normalize the currents such that the incoming current is one and obtain for the total transmission 
\be 
T = \sum_{n} T_n =  \sum_{n} |t_n|^2 ~\frac{\sin k_n}{\sin k_0}.
\ee
Finally, imposing current conservation ($J_- = J_+$) and considering relation (\ref{rtn}) we obtain
\be 
T=Re{E_0} = \sum_{n} |E_n|^2 ~\frac{\sin(k_n)}{\sin(k_0)} . 
\ee

\section{Location of the $T=0$ resonance}\label{App_B}
For very small barrier amplitude $\mu$ the resonances are located almost exactly at $\omega_{\pm}=2J \pm \epsilon$, just {\it before} the $n=\mp1$ chain enters the continuum. Therefore, we consider no additional unbound { channel}  (besides the $n=0$) on the corresponding side of the ladder. Remember that the recurrence relations is given by
\be
\gamma_n E_{n}  = E_{n-1}+E_{n+1}
\ee
with
\be
\gamma_n = \frac{2}{\mu} \left[\mp(1-\beta)\sqrt{(\epsilon + n\omega)^2-4J^2}-\beta(\epsilon\pm n\omega)\right]
\ee
for positive/negative $n$. For $\epsilon>0$ and $\omega>\omega_{+}$ ($\omega>\omega_{-}$), all coefficients $\gamma_{n\neq0}$ are real for $n<0$ ($n>0$). In the case of a zero transmission resonance we have $Re{E_0}=0$ which implies $Im{E_0}=0$ as well.
Thus, at the resonance we have $E_0=0$ such that all coefficients are determined by $E_{\pm1}$:
\bea
E_{\pm2}&=&\gamma_{\pm1} E_{\pm1}\\
E_{\pm3} &=& \gamma_{\pm2} E_{\pm2} -E_{\pm1} = (\gamma_{\pm1} \gamma_{\pm2} -1) E_{\pm1}\\
E_{\pm4} &=& \gamma_{\pm3} E_{\pm3} - E_{\pm2} = (\gamma_{\pm3} (\gamma_{\pm1} \gamma_{\pm2} -1)-\gamma_{\pm1})E_{\pm1}\\
E_{\pm5} &=& \gamma_{\pm4} E_{\pm4} -E_{\pm3} = ((\gamma_{\pm3} \gamma_{\pm4} -1)(\gamma_{\pm1} \gamma_{\pm2} -1)- \gamma_{\pm4}\gamma_{\pm1})E_{\pm1}
\eea
and so on. This series must converge, which in general gives a non-linear highly non-trivial condition on $\mu$, $\omega$ and $\epsilon$. However, note that all coefficients $E_n$ for larger $n$ contain terms that are proportional to $\gamma_{\pm1}$ or $(\gamma_{\pm1} \gamma_{\pm2} -1)$. All other factors like $(\gamma_{\pm3} \gamma_{\pm4} -1)$ are very large, so the only way that this series converges quickly is that
\be
\gamma_{\pm1} \gamma_{\pm2} -1 \ll \gamma_{\pm1} \ll 1
\ee
which yields $\gamma_{\pm1} \approx \gamma_{\pm2}^{-1}$ ($E_{\pm3}=0$). Therefore
\be
 \!\!\!\!\!\!\!\!\!\!\!\!\!\!\!\!\!\!\!\!\!\!\!\!\!\!\!\!\!\!\!\!\!\!\!\!\!\!\!\!\!\!\!\!\!\!\!\!\frac{2}{\mu} \left[\mp(1-\beta)\sqrt{(\epsilon\pm\omega)^2-4J^2}-\beta(\epsilon\pm\omega)\right]\approx\frac{\mu}{2} \left[\mp(1-\beta)\sqrt{(\epsilon\pm2\omega)^2-4J^2}-\beta(\epsilon\pm2\omega)\right]^{-1},
\ee
which is precisely equation (\ref{res_app}).

\section*{References}
\bibliographystyle{iopart-num}

\end{document}